# Semiconductor - Ferromagnetic Insulator - Superconductor Nanowires: Stray Field and Exchange Field

Yu Liu[1,2], Saulius Vaitiekėnas[2,3], Sara Martí-Sánchez[4], Christian Koch[4], Sean Hart[5,6], Zheng Cui[5,7], Thomas Kanne[2], Sabbir A. Khan[1,2], Rawa Tanta[1,2], Shivendra Upadhyay[3], Martin Espiñeira Cachaza[1,2], Charles M. Marcus[2,3], Jordi Arbiol[4,8], Kathryn A. Moler[5,6,7], Peter Krogstrup[1,2,*]

[1]Microsoft Quantum Materials Lab Copenhagen, 2800 Lyngby, Denmark.

[2]Center for Quantum Devices, Niels Bohr Institute, University of Copenhagen, 2100 Copenhagen, Denmark.

[3]Microsoft Quantum Lab Copenhagen, Niels Bohr Institute, University of Copenhagen, 2100 Copenhagen, Denmark.

[4]Catalan Institute of Nanoscience and Nanotechnology (ICN2), CSIC and BIST, Campus UAB, Bellaterra, 08193 Barcelona, Catalonia, Spain.

[5]Stanford Institute for Materials and Energy Sciences, SLAC National Accelerator Laboratory, Menlo Park, California 94025, USA.

[6]Department of Physics, Stanford University, Stanford, California 94305, USA.

[7]Department of Applied Physics, Stanford University, Stanford, California 94305, USA.

[8]ICREA, Pg. Lluís Companys 23, 08010 Barcelona, Catalonia, Spain.

Nanowires can serve as flexible substrates for hybrid epitaxial growth on selected facets, allowing for design of heterostructures with complex material combinations and geometries. In this work we report on hybrid epitaxy of semiconductor - ferromagnetic insulator - superconductor (InAs/EuS/Al) nanowire heterostructures. We study the crystal growth and complex epitaxial matching of wurtzite InAs / rock-salt EuS interfaces as well as rock-salt EuS / face-centered cubic Al interfaces. Because of the magnetic anisotropy originating from the nanowire shape, the magnetic structure of the EuS phase are easily tuned into single magnetic domains. This effect efficiently ejects the stray field lines along the nanowires. With tunnel spectroscopy measurements of the density of states, we show the material has a hard induced superconducting gap, and magnetic hysteretic evolution which indicates that the magnetic exchange fields are not negligible. These hybrid nanowires fulfil key material requirements for serving as a platform for spin-based quantum applications, such as scalable topological quantum computing.

Materials combining ferromagnetism and semiconductivity have been a long-standing goal for spin-based electronics.[1,2] A main challenge is related to the fact that most ferromagnets are metallic and therefore not tunable. Inducing magnetism from ferromagnetic insulators (FMI) could overcome this challenge. However, the device performance and reliability will depend on the band alignment and not the least the detailed structural quality. Especially, eliminating crystal defects and impurities plays a critical role in the development of spin-based quantum applications where reliable lifting of the spin degeneracies is key to technology realization. Topological materials are a class of materials which hold promise in quantum information processing as a fault tolerant basis for operations of topologically protected quasiparticles[3-6]. Due to its ultra-pure environment, molecular beam epitaxy (MBE) has been the preferred synthesis method for such materials with examples like high mobility two-dimensional semiconductors (GaAs/AlGaAs[7] and HgTe/HgCdTe[8]), that have revealed quantum Hall states (fractional or spin), topological insulator-superconductor heterostructures (nonmagnetic $Bi_2Te_3/NbSe_2$[9,10], and ferromagnetic $(Cr_xBi_ySb_{1-x-y})_2Te_3/Nb$[11,12]) which have shown states of different topological order. Another class of topological states is the so-called Majorana bound states (MBS), which are predicted to exist at the ends of one-dimensional *p-wave* superconductors.[13,14] Recently, signatures of MBS have been reported in various platforms: hybrid III-V semiconductor-superconductor (SE/SU) nanowires (NWs)[15-23], ferromagnetic atomic Fe chains on Pb superconducting surface[24-27], iron-based superconductors[28-30] and most recently carbon nanotubes with stray field induced magnetism[31].

However, obtaining signatures of topological states is only the first step towards realization of topological protected states and applications based thereon. Reproducibility and control of the fabrication processes on the atomic level will be needed for an effective optimization process towards topological protection. Another concern for large scale operations comes from needs of external magnetic fields which are used to lift the spin degeneracies in SE/SU NWs. Therefore, it is of interest to integrate materials that are topological without the need for external applied fields. Composite materials using FMIs in close proximity to a SE/SU structure have been proposed as a solution to reach a zero-field topological state[32], where the effective Zeeman splitting is induced by an magnetic exchange coupling with the FMI (see for example see Ref. 33). However, the uniformity of this exchange field will be critical to form an extended topological phase, which sets requirements to the quality of the interface between the FMIs and the SE (and/or SU). The ideal interface will always be an impurity free and fully coherent epitaxial interface. Moreover, a uniform morphology is another requirement for the FMI phase as magnetic stray fields from disordered films will penetrate the NWs causing fluctuations in the band structure.

There have recently been reports on significant magnetic exchange coupling from EuS in materials such as graphene and $Bi_2Se_3$[34,35], spin-polarized tunneling in Au/EuS/Al[36,37] and splitting of the superconducting density of states in EuS/Al films[38]. From a structural point of view, InAs/EuS hybrid structure seems like an ideal hybrid for a tunable spin-lifted semiconducting material with strong spin-orbit coupling. As single crystal EuS epitaxy can be obtained at temperatures that are compatible with InAs stability[39], and because EuS has a cubic rock-salt structure with a small bulk lattice mismatch to the cubic InAs zinc-blende phase of ~1%, it allows for fully coherent InAs/EuS epitaxy[40]. Because of the temperature restriction on the III-V stability, fabrication of III-V/FMI systems, Eu-based chalcogenides[41,42] seem to have larger potential than other FMIs reported in literature, such as ferrites[43,44], garnets[45] and layered Cr-based trihalides[46,47]. Here, we report on the tri-crystal epitaxy of InAs/EuS/Al 'SE-FMI-SU' nanowires and show these uniquely matched materials turns out to fulfill key requirements for applications in fields such as spintronics and topological quantum computing.

Free standing semiconducting nanowires can be thought as one-dimensional 'substrates' that allows hybrid epitaxial growth on selected facets from all radial directions. This flexibility opens for a variety of advanced design possibilities of heterostructures that could lead to quantum wires with novel properties. One interesting path involves epitaxy of both a superconductor and a ferromagnetic insulator on the semiconductor NW facets, due to the potential of opening a topologically protected gap without external magnetic field[32]. Obviously, many detailed requirements for the morphology, crystal structure and composition need to be fulfilled for realizing such an isolated hybrid quantum state. To realize high quality interfaces and well defined crystal orientations we performed tri-crystal epitaxy of InAs/EuS/Al NWs where the InAs NWs is grown vertically on the substrate via the Vapor-Liquid-Solid method[48] and the subsequent growth of EuS and Al on the InAs NW facets are carried out in-situ without breaking the ultra-high vacuum (see methods for detail). The InAs/EuS/Al NWs are approximately 10 µm long and 100 nm in diameter (Figure 1a), the EuS of ~4.5 nm was grown on two of the six dominant {11-20}-type side facets followed by ~6 nm of Al deposited on two side facets overlapping one side facet with EuS. The area shown with the cross-section atom-resolved high-angle annular dark-field (HAADF) aberration corrected scanning transmission electron microscope (AC-STEM) image in Figure 1b corresponds to one of the {1-100}-type facets. The high-resolution transmission electron microscopy (HRTEM) image along NW growth direction in Figure 1c further confirms the domain match in Figure 1b. The EuS structure shows no rotation in this orientation, while a more complex hybrid structure appears on the {11-20}-type side facets, which are presented in Supplementary Information Figure S1. The corresponding indexed power spectrum in Figure S2 shows the tri-crystal epitaxial relationship where the zone axis of InAs [1-100] // EuS [1-10] // Al [1-10] are aligned. The atomic positions are highlighted by superimposed atomic columns on both InAs/EuS and EuS/Al interface in Figure 1d. The corresponding atomic position simulation on a top view in Figure 1e shows a remarkable epitaxial relationship between rock-salt EuS and wurtzite InAs $(3_{[332]}/4_{[000\text{-}1]}, 0.0\%)_{//} \times (1_{[1\text{-}10]}/1_{[11\text{-}20]}, -1.5\%)_{\perp}$ (the notation method follows the same definition given in the previous report[49]). Therefore, in terms of bi-crystallography, given only the lowest free energy match is formed, the EuS grows without grain boundaries (see Figure S3). The top view of the interfacial atomic positions, using relaxed bulk lattice parameters, shows the epitaxial relationship of Al on EuS $(4_{[-1-1-1]}/3_{[332]}, 0.2\%)_{//} \times (3_{[1-10]}/2_{[1-10]}, 1.8\%)_{\perp}$ in Figure 1f. The bi-crystal variant degeneracy is 2 while the order of PRS is 1, which means there should be two degenerated Al orientations.

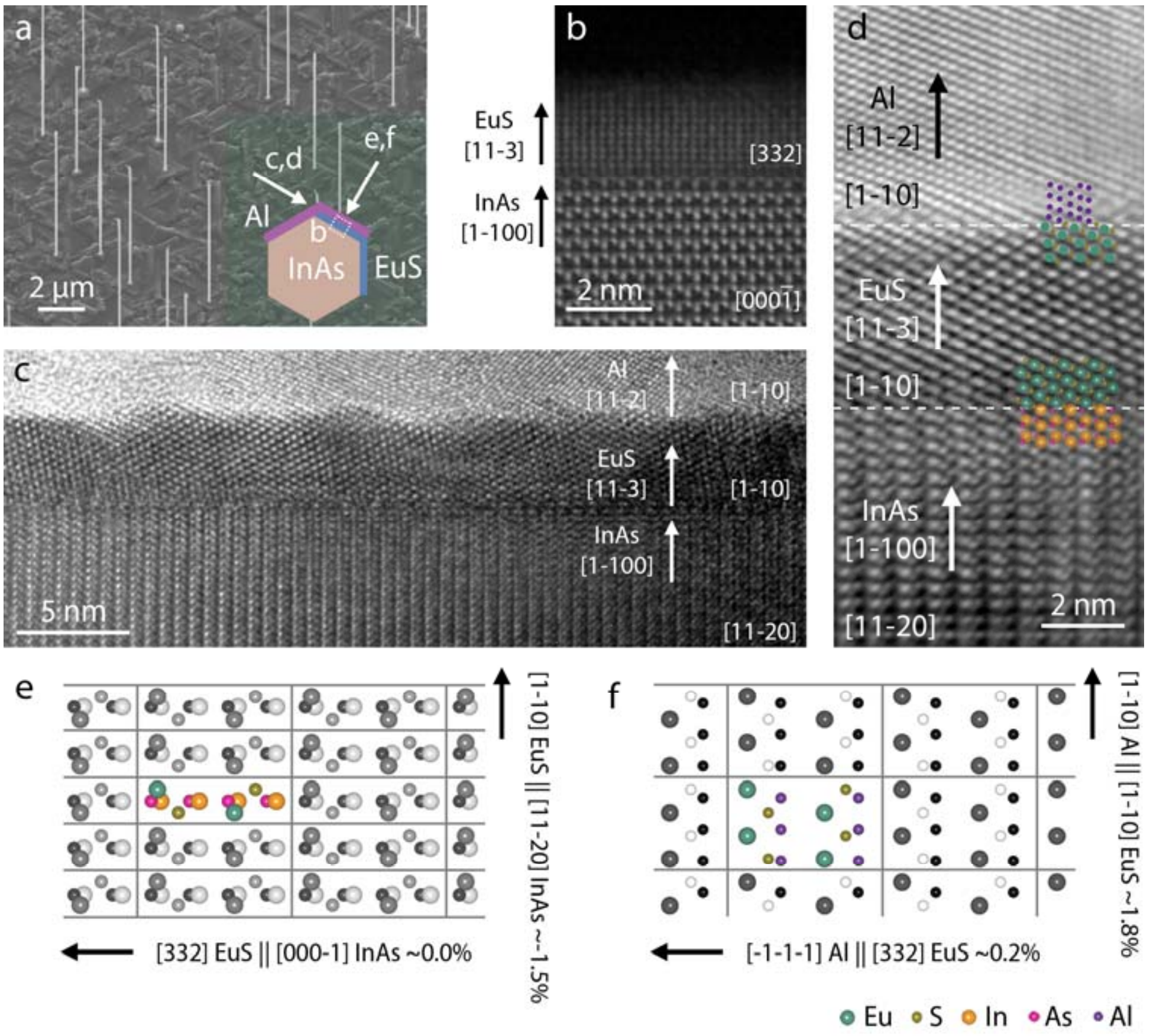

**Figure 1 | InAs/EuS/Al VLS NW. a,** Scanning electron microscope (SEM) image (viewing angle 30° from normal) of InAs/EuS/Al hybrid NWs grown vertically on the substrate. The schematic inset indicates the view directions in the following sub figures. **b,** HAADF STEM cross-sectional micrograph obtained from the {1-100}-type facet of the NW. **c,** HRTEM image obtained with zone axis transverse to the {1-100}-type facet of the InAs/EuS/Al NW. The crystal directions of InAs, EuS and Al are marked with arrows in **b** and **c**. **d,** The HRTEM image to show atomic arrangement near the interface. The atomic positions are suggested by superimposing atomic columns. **e,f,** The top view of the interfacial atomic positions, [11-3] EuS vs. [1-100] InAs and [11-2] Al vs. [11-3] EuS, to show lattice match, using the lattice constants taken from bulk face-centered cubic Al, rock-salt EuS, and wurtzite InAs. Grey solid lines indicate primitive domains. Vectors show the parallel and transverse directions including the corresponding residual mismatch.

To examine the magnetic stray fields of VLS grown InAs/EuS/Al NWs we probe the field distribution using a scanning Superconducting QUantum Interference Device (SQUID) measurement setup[50] from which we extract detailed information about the ferromagnetic domain structure. The field lines of a single ferromagnetic domain aligns parallel to the VLS NW axis are illustrated in the inset of Figure 2a. The illustration is overlaid with a SQUID measurement of a 10 μm long VLS grown InAs NW to demonstrate the correspondence between the magnetic field and the SQUID image. The NW is grown with ~4.5 nm EuS and ~6 nm Al on facets as described in Figure 1. The NW magnetometry at 5 K demonstrates the feature of a single magnetic domain (Figure 2a), taking into account the shape of the SQUID probe itself (the inset of Figure 2a). The SQUID mapping was further executed on the whole area with transferred NWs cooled to 5 K with a minor external field of 65 Oe for saturation. In the mapping, each magnetic dipole can be matched to a single NW in Figure 2b (more details are shown in Figure S4 and the corresponding peak-to-peak magnetometry signal is listed in Table S1). It is also found that the magnetization directions are always along NW axis. The DC scanning result (Figure S5) indicates the Curie temperature at 19 K, slightly higher than that of bulk EuS (16-17 K).

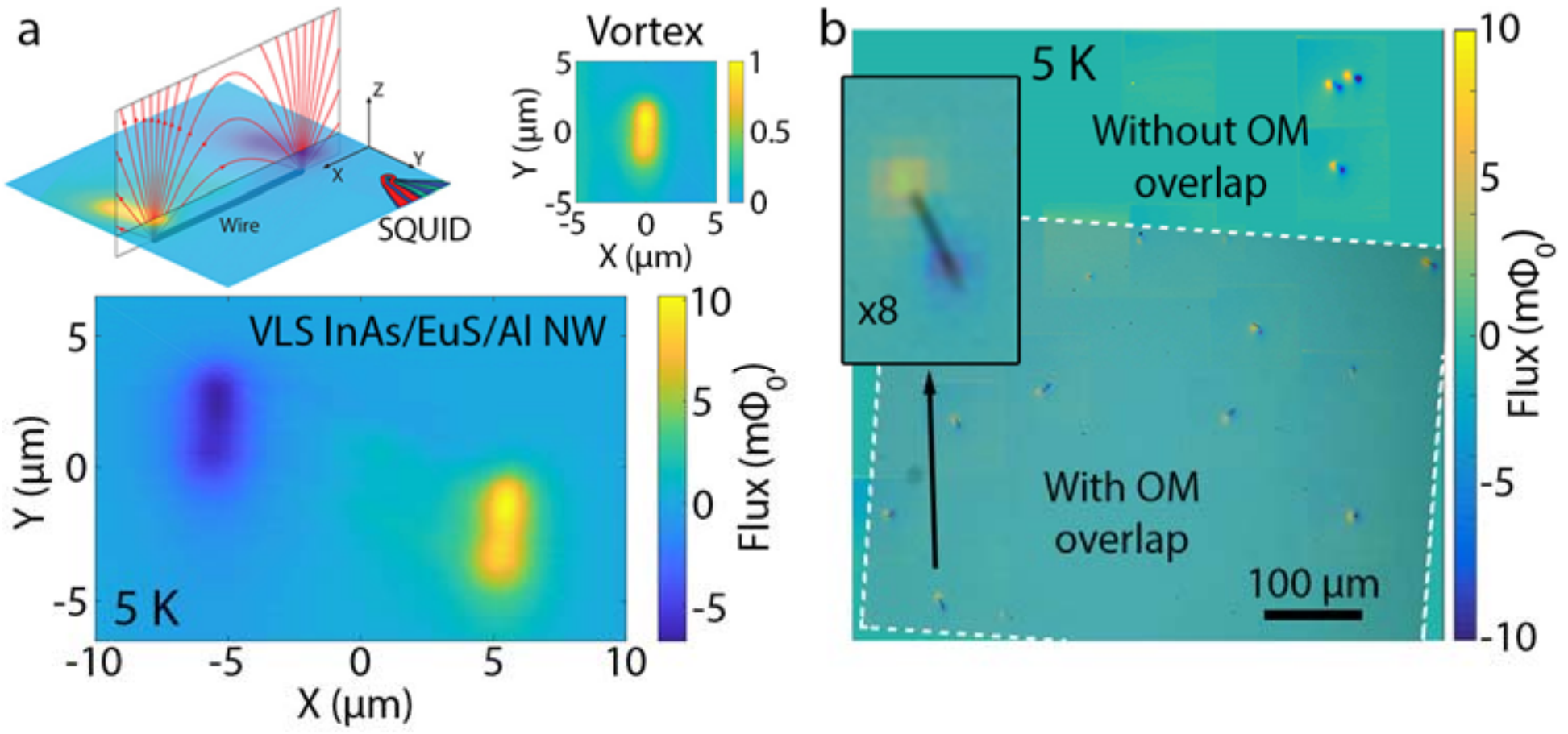

**Figure 2 | Magnetization of InAs/EuS/Al VLS NWs. a,** SQUID measurement of a 10 μm long InAs/EuS/Al VLS NW at 5 K. The SQUID measure an out-of-plane magnetic flux at both ends of the wire, shown in blue and yellow. Inset schematics show the field lines from the dipole, how they are picked up by the SQUID, and the SQUID Point-Spread-Function as measured by imaging a superconducting vortex. **b,** The SQUID mapping on the area having randomly oriented InAs/EuS/Al VLS NWs. The one-to-one relation is confirmed between the measured dipoles and the NWs, among which one NW is zoomed in. The edge of the overlapping optical microscope image is marked with dashed lines.

To avoid variations in the potential along the NWs, it is important to minimize the contribution of the magnetic stray field of EuS penetrating into the NWs. The external magnetic field is simulated based on the current InAs/EuS/Al NW structure. For details about the field simulation and the distribution of the total magnetic field and the magnetic field along [0001] inside the InAs NW, see Supplementary Information and Figure S7. The single domain simulation shows that most of the magnetic field are distributed at the dipole and only minor stray field penetrates the NWs. This is appropriate for future applications of single domain NWs for which the corresponding Zeeman lifting are based on exchange coupling to the FMI.

To investigate the superconducting and ferromagnetic proximity effects, we fabricated a tunnelling-spectroscopy device using VLS NWs for low-temperature transport measurements. The devices were fabricated as shown in Figure 3a using standard electron-beam lithography techniques. The devices are equipped with ohmic contacts for transport and side-gate electrodes for potential tuning. Differential tunnelling conductance, $G = dI/dV$, is a direct measure of local density of states at the end of the wire. Cutter gate at voltage $V_C$ is used to tune the tunnelling barrier. Upper and lower gates at voltages $V_U$ and $V_D$ are used to tune the electronic wave function in the wire. The cross section of the wire with the relative shell orientation is illustrated in Figure 3b. In order to detect the superconducting and ferromagnetic proximity effects respectively, the wires are grown without an overlap between Al and EuS. In the tunnelling regime, the wire displays a superconducting gap of $\Delta \sim 230$ μeV, inferred from the broad coherence peaks, see Figure 3c. The subgap conductance is suppressed by two orders of magnitude with respect to the normal state conductance indicating a good, unpoisoned proximity effect. Figure 3d shows the evolution of the wire density of states as the magnetic field, $\mu_0 H$, was swept from 0.2 T to -0.2 T. The coherence peaks are split and rounded at $\mu_0 H = 0.2$ T. The splitting reduces upon decreasing the magnetic field. As the field passes zero, the splitting keeps reducing and reaches minimum around -30 mT, where the gap is maximal. Ramping the field further to the negative direction starts to split peaks again. At $\mu_0 H = -0.2$ T the spectrum is indistinguishable from the one at $\mu_0 H = 0.2$ T. Qualitatively similar phenomenology was observed by sweeping the field from negative to positive magnetic field value, except the splitting was minimal on the opposite side of $\mu_0 H = 0$. We illustrate this hysteretic behaviour of the tunnelling spectrum by subtracting two conductance maps, measured while sweeping the magnetic field two opposite directions, see Figure 3e. It is clear that the two spectra are identical above $|\mu_0 H| \sim 0.1$ T and have maximal difference around ± 30 mT. Because the stray fields observed by scanning squid measurements are negligible and the device fabrication does not remove EuS (see Figure S8), we interpret that these results as arising from the ferromagnetic EuS induced exchange field, with a coercive field of ~30 mT, where the magnetisation is zero. The coherence peaks at zero field are broad presumably due to the residual exchange field induced splitting, that is smeared out by the strong spin-orbit coupling in the wire.

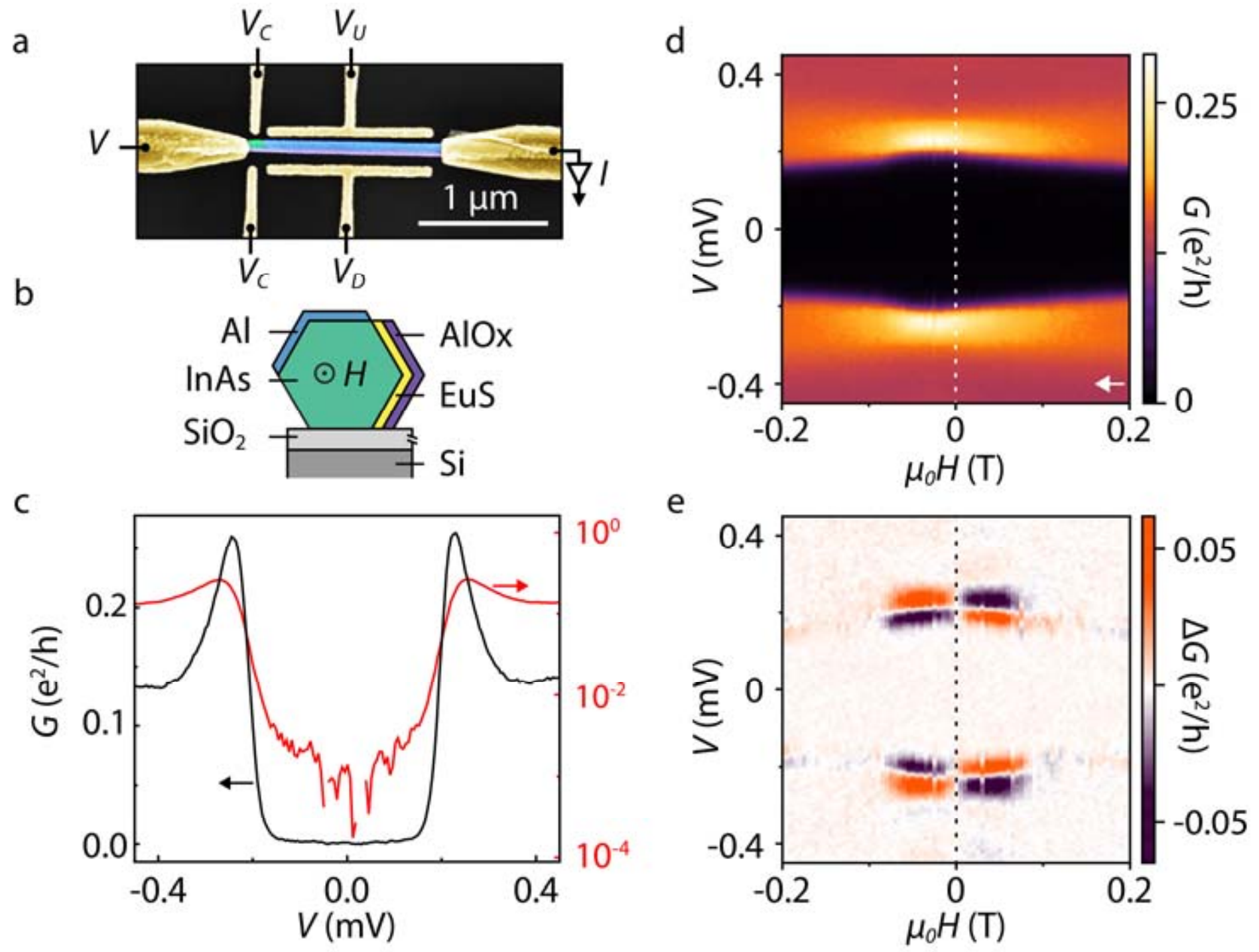

**Figure 3 | Tunneling spectroscopy. a,** False-color electron micrograph of the tunnelling device, showing InAs nanowire (green), two-facet Al shell (blue), EuS coated with AlOx (purple) as well as Ti/Au contacts and side-gates (golden). **b,** Schematic device cross section, showing the orientation of external magnetic field strength, H, and Al as well as EuS shells relative to the bottom-gate. Note that the wires studied here have no overlap between EuS and Al shells, which is different from those mentioned above. **c,** Differential conductance, G, measured as a function of applied source-drain bias voltage, V, in linear (black, left) and logarithmic (red, right) scales. The spectrum shows a hard induced superconducting gap with broad coherence peaks ~ 230 μeV. **d,** Evolution of the spectrum shown in **c** with magnetic field, $\mu_0H$. The conductance map was recorded while ramping the field from 0.2 T to -0.2 T, as indicated by the white arrow. The gap is maximal at around -30 mT, indicating hysteric wire behavior. **e,** Difference of two conductance maps, that were measured while sweeping the magnetic fields in opposite directions, further illustrating the hysteresis of the wire. The data were taken at $V_C$ = -6.65 V, $V_U$ = 2.6 V and $V_D$ = -1 V.

Selective area epitaxy of nanostructures[51,52] is a promising synthesis method for scalable applications. Here we grow EuS epitaxially on one facet of selective area grown (SAG) InAs NWs on (001) GaAs substrates (see Figure 4a,b) with the aim of forming single magnetic domains with shape controlled spin directions. A cross sectional STEM micrograph in Figure 4c shows that EuS is deposited on a (01-1) side facet of a [100] InAs NW. The atom-resolved image in Figure 4d shows the lattice matched (01-1)/(01-1) interface between the EuS and InAs phases, as also seen in the planar (100) platforms[40]. The relevant index power spectrum is shown in Figure S9. The atomic species are analyzed based on the chemical sensitivity (Z contrast) of the HAADF STEM[53,54] and thus shown with atomic columns superimposed on the image. In this case, no compression can be observed in GPA (Figure 4e and more details in Figure S10 in the Supplementary Information). Interestingly, this is in strong contrast to the (001)/(001) cubic-cubic combination on planar samples as recently reported[40]. The in-plane dilatation and rotation maps on (022) planes also demonstrate interface with no misfit dislocation. The interface is simulated in Figure 4f. The order of PRS of EuS on InAs is 1 and its bi-crystal variant degeneracy is 1.

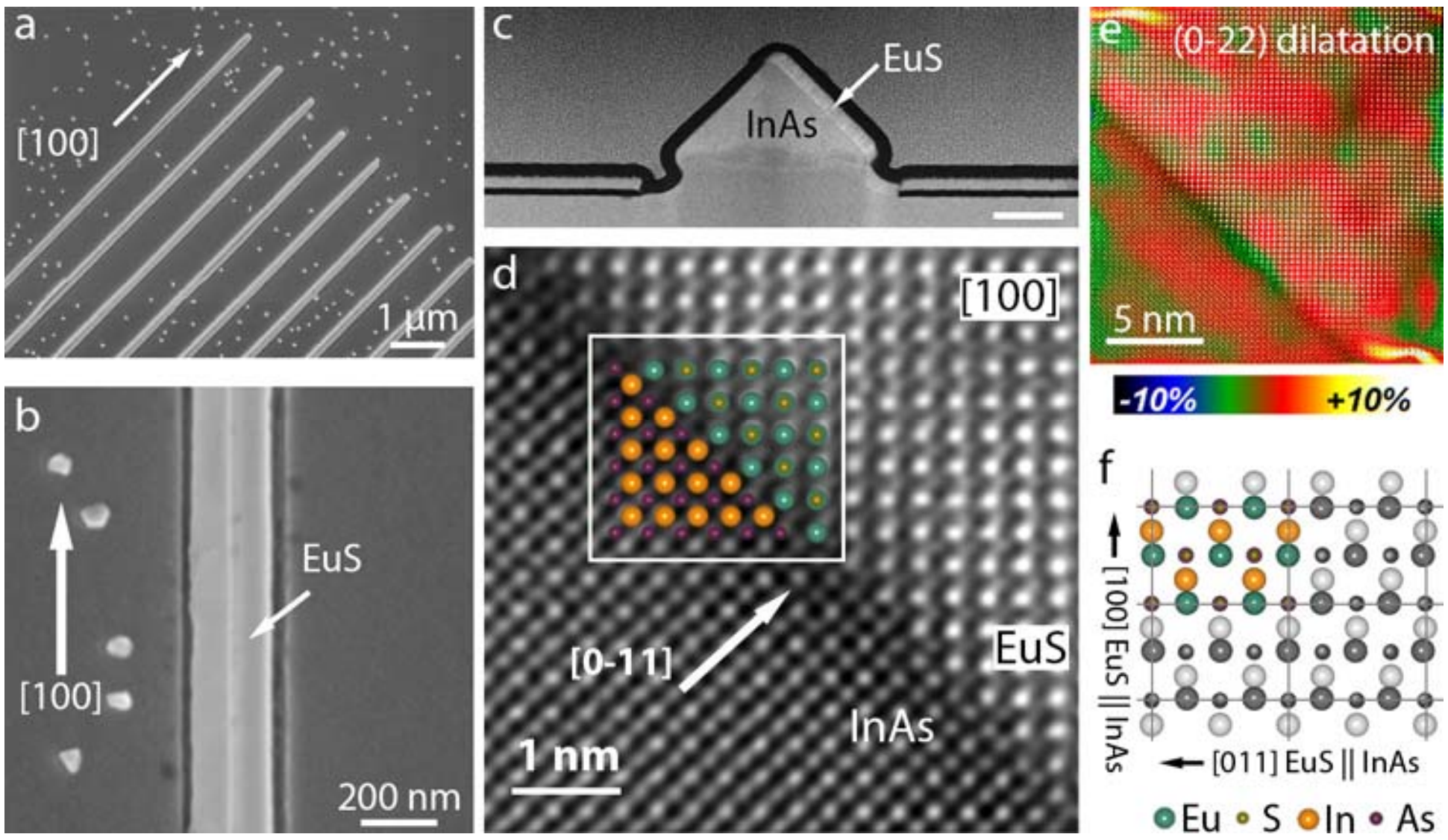

**Figure 4 | InAs/EuS SAG NW. a,** A top-view SEM image of [100] InAs SAG NW array with EuS on one of the side facets. **b,** A top-view SEM image with high magnification of a single [100] InAs/EuS SAG NW. **c,d,** A HAADF AC-STEM cross-sectional micrograph of a SAG InAs/EuS NW. The simulated model of the atomic positions near the interface is based on the mass-dependent intensity in the HAADF AC-STEM using the bulk component intensities as reference. The scale bar in **c** represents 100 nm. **e,** The GPA dilatation map on (0-22) planes shows coherent epitaxial match without any apparent compression. The map overlaps with the corresponding cross-sectional image to show the position of the interface. **f,** Top view of the interfacial atomic positions, [0-11] EuS vs. [0-11] InAs, to show lattice match, using the lattice constants taken from rock-salt EuS and zinc-blende InAs. Grey lines indicate primitive interfacial domains.

We find that the intrinsic magnetization direction of the EuS when grown in the VLS NWs geometry are always along NW axis. This is attributed to the shape anisotropy like a single domain nano-bar-magnet. The reason is that magnetization perpendicular to NW will increase the free energy of the NW system. However, it is possible to manipulate magnetization to non-parallel directions of NWs by growing thin EuS layers between the NWs. Figure 5a shows the scanning SQUID result of a single SAG NW along InAs [100]. Different from VLS NWs, it only needs ~3 Oe to saturate magnetization. A field perpendicular to the NW was applied to this NW, but the direction of its magnetization cannot be fully aligned perpendicular to the NW (Figure S11). In contrast, we observe apparent perpendicular magnetization in an array of SAG NW grown along InAs [110], as shown in Figure 5b. It is also possible for an InAs [100] NW array (Figure S12 in the supplementary information). So the NW array is helpful to obtain the magnetization perpendicular to NW axis, suggesting that the shape anisotropy rather than NW crystal orientations decides the possibility to manipulate the magnetization orientation.

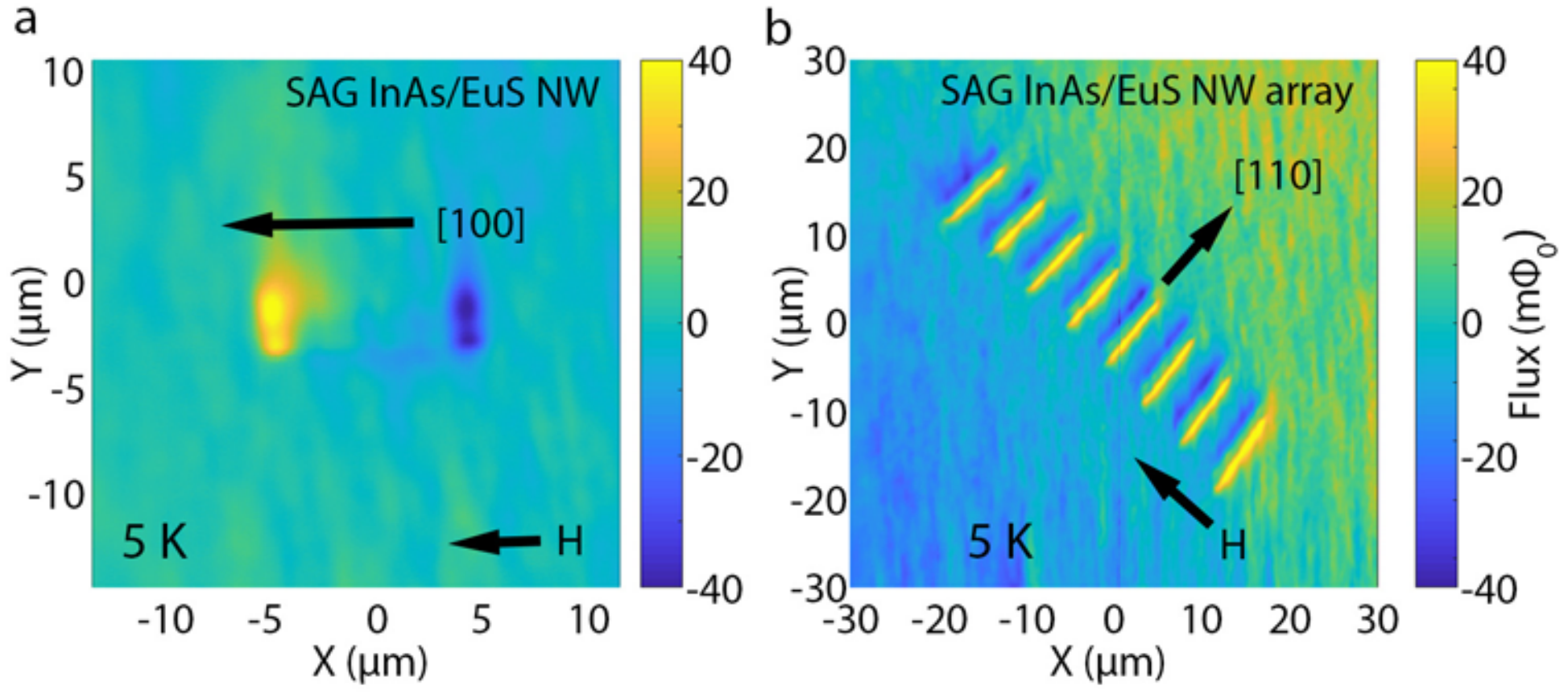

**Figure 5 | Magnetization of InAs/EuS SAG NWs. a,** SQUID measurement of a SAG InAs/EuS [100] NW at 5 K, showing the feature similar to that of VLS NWs if cooling with magnetic field along [100] (parallel to the NW axis). **b,** The SQUID mapping of SAG InAs/EuS [110] NW array at 5 K with in-plane magnetic field cooling perpendicular to [110], showing the possibility to manipulate the magnetization direction of InAs/EuS NWs.

---------------------------------------------------

In summary, we present semiconductor-ferromagnetic insulator-superconductor (InAs/EuS/Al) quantum nanowires with epitaxial interfaces. The epitaxy of both cubic EuS and Al on the side facets of hexagonal InAs NW crystals, resulting in complex tri-crystal epitaxial match, with in-plane epitaxial relationship. We examined the magnetic stray field distribution of these NWs and show single magnetic domains with a Curie temperature of 19 K along the NW axis dictated by shape anisotropy. We show that the InAs nanowires have a hard induced superconducting gap and show a magnetic hysteretic evolution of the density of states which presumably stems from the exchange fields of EuS. We further show the possibility to obtain EuS epitaxy on selective area grown InAs nanowires, whose magnetization orientation can be manipulated. These results show that the NW tricrystals (InAs/EuS/Al) fulfill key requirements for realization of topological superconducting NWs without external fields.

**Methods:** Au-catalyzed InAs NWs were grown on 1/4 2-inch undoped n-type (111)B zinc-blende InAs wafers (Semiconductor Wafer Inc.) with the VLS method in III-V growth chamber of a solid-source Varian GEN-II MBE system with the background pressure below $10^{-10}$ Torr.[49] The natural oxidized layers were desorbed from the InAs by heating the substrate to a temperature of 525 °C (measured with a pyrometer) under As2 overpressure protection for 360 s. The Au beam was shot to the substrate in about 1 sec as catalysis before NW growth. InAs NWs were grown with a substrate temperature at 447 °C (with a thermo-coupling back sensor) for 2100 s. The corresponding planar growth rate is 0.5 μm/hr and the As2/In ratio of 28. Further InAs radial growth was employed to control NW diameters with a substrate temperature at 350 °C (with a thermo-coupling back sensor) for 300 s with the As2/In ratio of 22.5. The substrates with as-grown InAs NWs were transferred from the III-V growth chamber to the metallic chamber right after growth. The substrates had about 45° tilt towards the EuS source. ~4.5 nm of EuS epitaxy was grown on two side facets of NWs with in-situ electron beam evaporation with a substrate temperature of 180 °C and an average growth rate of 0.02 nm/s. To further prepare InAs/EuS/Al NWs, the substrates with InAs/EuS NWs were transferred from the metallic chamber back to the III-V growth chamber after EuS deposition. The Al deposition temperature is below -30 °C and the deposition rate is 0.02 nm/s. The Al thickness is ~6 nm. The growth of SAG InAs was the same as that in the previous report[51]. The growth of EuS on SAG InAs employed the same setting of the growth of EuS layers on InAs NWs mentioned above but its thickness is around 20 nm. The deposition direction is perpendicular to [100] NWs. AlOx was deposited to protect the EuS layers after EuS deposition in the same chamber for the transport measurements.

Structural characterization was performed on the samples by means of atomic resolution HAADF AC-STEM in a Titan FEI microscope. The cross-section lamellae have been obtained by Focused Ion Beam HELIOS 600 FIB cutting perpendicular to the growth direction of the selected NWs. HRTEM was carried out on a JEOL JEM 3000F and a FEI F20, both with field emission guns. In order to map the deformation present at the NWs, we have used geometric phase analyses (GPA), developed by Hytch.[55,56] Structural simulation of the NW crystals in Figures. 1d (InAs/EuS interface),1e, 1f, 4d, 4f and S1b were performed using VESTA[57].

NWs were imaged using a scanning SQUID susceptometer[50]. The SQUID measures the magnetic flux through an integrated pickup loop while it scans over the sample surface. A concentric field coil can carry a current to locally apply a magnetic field. Superconducting shields and gradiometric design reduce the contribution of spurious magnetic field sources. The sensors operate in two main modes: magnetometry and susceptometry. Magnetometry images show the DC flux through the pickup loop as a function of its positions. Susceptometry images show a lock-in measurement of the AC flux through the pickup loop when there is an AC current through the field coil. The gradiometric design includes an astatically wound pickup loop / field coil pair that cancels most of the applied flux so the contribution of the sample's response is measurable in the susceptometry signal. In this work, the sensors only operated in the magnetometry mode.

Two sensors with different overall dimensions were employed to generate the images presented here (see Ref. 50). Generally, a smaller pickup loop size affords superior spatial resolution at the expense of sensitivity; the details of this relationship depend on the exact nature of the magnetic field source. The smaller of the two sensors had a pickup loop inner diameter of 200 nm, and a field coil inner diameter of 1 micron. These sensor dimensions are relevant for data depicted in Figures. 2a, 5a, 5b, S11 and S12. For these measurements, the sensor was located at a nominal distance of 700 nm above the substrate. The larger sensor had a pickup loop inner diameter of 6 microns, and a field coil inner diameter of 12 microns. These sensor dimensions are relevant for data depicted in Figures. 2b, S4, S5 and Table S1. For these measurements, the sensor was positioned nominally 1 micron above the substrate.

The measured VLS nanowire device was fabricated using standard e-beam lithography techniques on a degenerately n-doped Si substrate with a thermal $SiO_x$ (200 nm) capping. The Al shell was selectively wet etched in MF-321 photoresist developer for 90 s, while the EuS shell was protected by the $AlO_x$ coating. Normal Ti/Au (5/150 nm) ohmic leads were metalized after in-situ Ar-ion milling (RF-ion source, 15 W, 18 mTorr, 5 min), followed by deposition of Ti/Au (5/100 nm) side gates. Transport measurements were conducted with standard ac lock-in techniques in a dilution refrigerator with a base temperature of 20 mK and a three-axis vector magnet.

**ASSOCIATED CONTENT**

**Supporting Information**

The Supporting Information is available free of charge on the ACS Publications website at DOI: Structural information of InAs/EuS/Al VLS NWs (Figure S1-S3), magnetic fields of InAs/EuS/Al VLS NWs (Figure S4- S8, Table S1), structural information of InAs/EuS SAG NWs (Figure S9,S10), stray field distribution of InAs/EuS SAG NWs (Figure S11,S12).

**AUTHOR INFORMATION**

**Corresponding Author**

*E-mail: krogstrup@nbi.dk.

**Note**

The authors declare no competing financial interests.

**ACKNOWLEDGMENTS**

We thank Mingtang Deng, Tomaš Stankevič, Chetan Nayak, Roman Lutchyn, Dmitry Pikulin for fruitful discussions, C. B. Sørensen for technical assistance, and Filip Křížek and Joachim Sestoft for the help on sample preparation. We acknowledge financial support from the Microsoft Quantum initiative, from the Danish Agency for Science and Innovation through DANSCATT, from the European Research Council under the European Union's Horizon 2020 research and innovation program (grant agreement n° 716655), and from the international training network 'INDEED' (grant agreement n° 722176). S.M.S. acknowledges funding from "Programa Internacional de Becas "la Caixa"-Severo Ochoa". S.M.S., C.K. and J.A. acknowledge funding from Generalitat de Catalunya 2017 SGR 327. ICN2 is supported by the Severo Ochoa program from Spanish MINECO (Grant No. SEV-2017-0706) and is funded by the CERCA Programme / Generalitat de Catalunya. Part of the present work has been performed in the framework of Universitat Autònoma de Barcelona Materials Science PhD program. Part of the HAADF-STEM microscopy was conducted in the Laboratorio de Microscopias Avanzadas at Instituto de Nanociencia de Aragon-Universidad de Zaragoza. ICN2 acknowledge support from CSIC Research Platform on Quantum Technologies PTI-001.

Supplementary information

# Semiconductor - Ferromagnetic Insulator - Superconductor Quantum Wires: Stray Field vs. Exchange Field

Yu Liu[1,2], Saulius Vaitiekėnas[2,3], Sara Martí-Sánchez[4], Christian Koch[4], Sean Hart[5,6], Zheng Cui[5,7], Thomas Kanne[2], Sabbir A. Khan[1,2], Rawa Tanta[1,2], Shivendra Upadhyay[3], Martin Espiñeira Cachaza[1,2], Charles M. Marcus[2,3], Jordi Arbiol[4,8], Kathryn A. Moler[5,6,7], Peter Krogstrup[1,2,*]

[1]Microsoft Quantum Materials Lab Copenhagen, 2800 Lyngby, Denmark.

[2]Center for Quantum Devices, Niels Bohr Institute, University of Copenhagen, 2100 Copenhagen, Denmark.

[3]Microsoft Quantum Lab Copenhagen, Niels Bohr Institute, University of Copenhagen, 2100 Copenhagen, Denmark.

[4]Catalan Institute of Nanoscience and Nanotechnology (ICN2), CSIC and BIST, Campus UAB, Bellaterra, 08193 Barcelona, Catalonia, Spain.

[5]Stanford Institute for Materials and Energy Sciences, SLAC National Accelerator Laboratory, Menlo Park, California 94025, USA.

[6]Department of Physics, Stanford University, Stanford, California 94305, USA.

[7]Department of Applied Physics, Stanford University, Stanford, California 94305, USA.

[8]ICREA, Pg. Lluís Companys 23, 08010 Barcelona, Catalonia, Spain.

## 1. Structural information of InAs/EuS/Al VLS NWs

1.1 The atomic arrangement of the InAs/EuS interface on {11-20}-type facets

The atomic arrangement of the InAs/EuS heterointerface after epitaxy on {11-20}-type facets is demonstrated with the atomic resolution HAADF AC-STEM view of the NW cross-section shown in Fig. S1a. The polycrystalline EuS shell is composed of ~4.5 nm high grains. Most of the crystal grains appear oriented along the [112] zone axis, which lays parallel to the [0001] axis of InAs. The in-plane epitaxial relationship is the (11-20) InAs with the (1-10) EuS planes, highlighted by superimposed atomic columns. However, in order to better accommodate the atomic positions of the Eu and S atoms and thus minimize the in-plane strain, the (1-10) EuS plane suffers up to ±15º rotation with respect to (11-20) InAs, creating grain boundaries when two different grains with opposite rotation angles interact. The observed average lateral size for the EuS grains is about 5 nm. The rotation largely reduces the mismatch from the domain match $(1_{[112]}/1_{[0001]}, 4.5\%)_{//} \times (2_{[11-1]}/3_{[1-100]}, -7.1\%)_{\perp}$ to $(1_{[112]}/1_{[0001]}, 4.5\%)_{//} \times (2_{15^{o}\ off\ [11-1]}/3_{[1-100]}, -3.8\%)_{\perp}$. The indexed power spectrum verifies the corresponding crystal orientation in Fig. S1b. The corresponding epitaxial relationship is suggested by atomic position simulation on a top view, but primitive domains are hard to identity due to the large lattice mismatch. This would only be possible if gaining more information about structural relaxation. A schematic of a single InAs NW with EuS epitaxy on two {11-20}-type side facets to help understanding the obtained samples is sketched accordingly (Fig. S1 inset on the left).

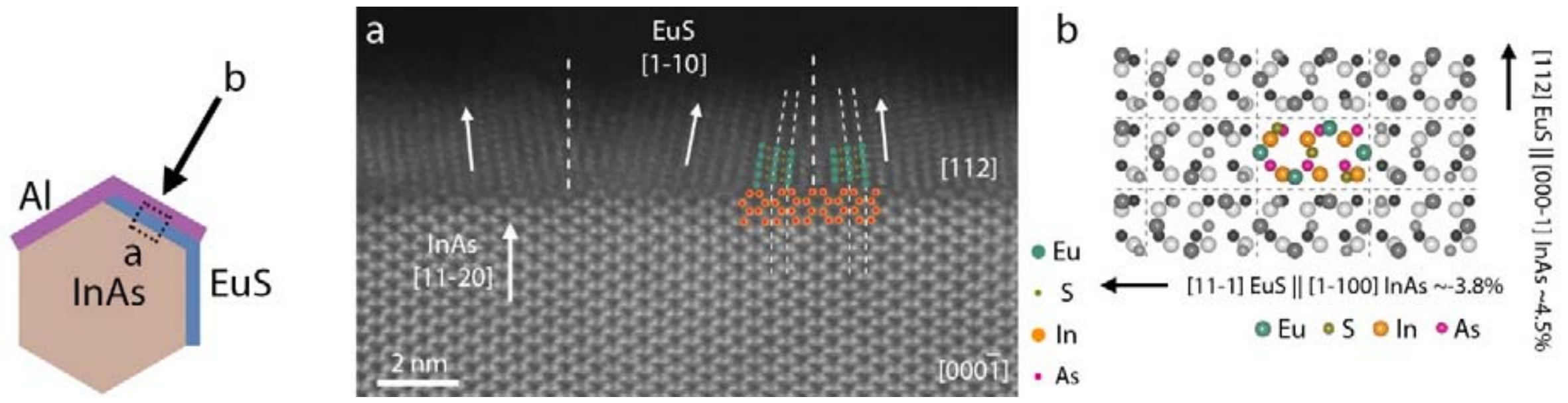

Figure S1 | a, HAADF STEM cross-sectional micrograph obtained in the {11-20} main facet of the InAs NW. The EuS grain boundaries in f are marked by dotted lines. b, The top view of the interfacial atomic positions, [1-10] EuS vs. [11-20] InAs, to show lattice match, using the lattice constants taken from bulk RS EuS and WZ InAs. Grey dashed lines only suggest primitive domains as reference because the lattice mismatch is large. Vectors show the parallel and transverse directions including the corresponding residual mismatch. The schematic on the left indicates the view directions.

1.2 Structural information of InAs/EuS/Al VLS NWs on {1-100}-type facets

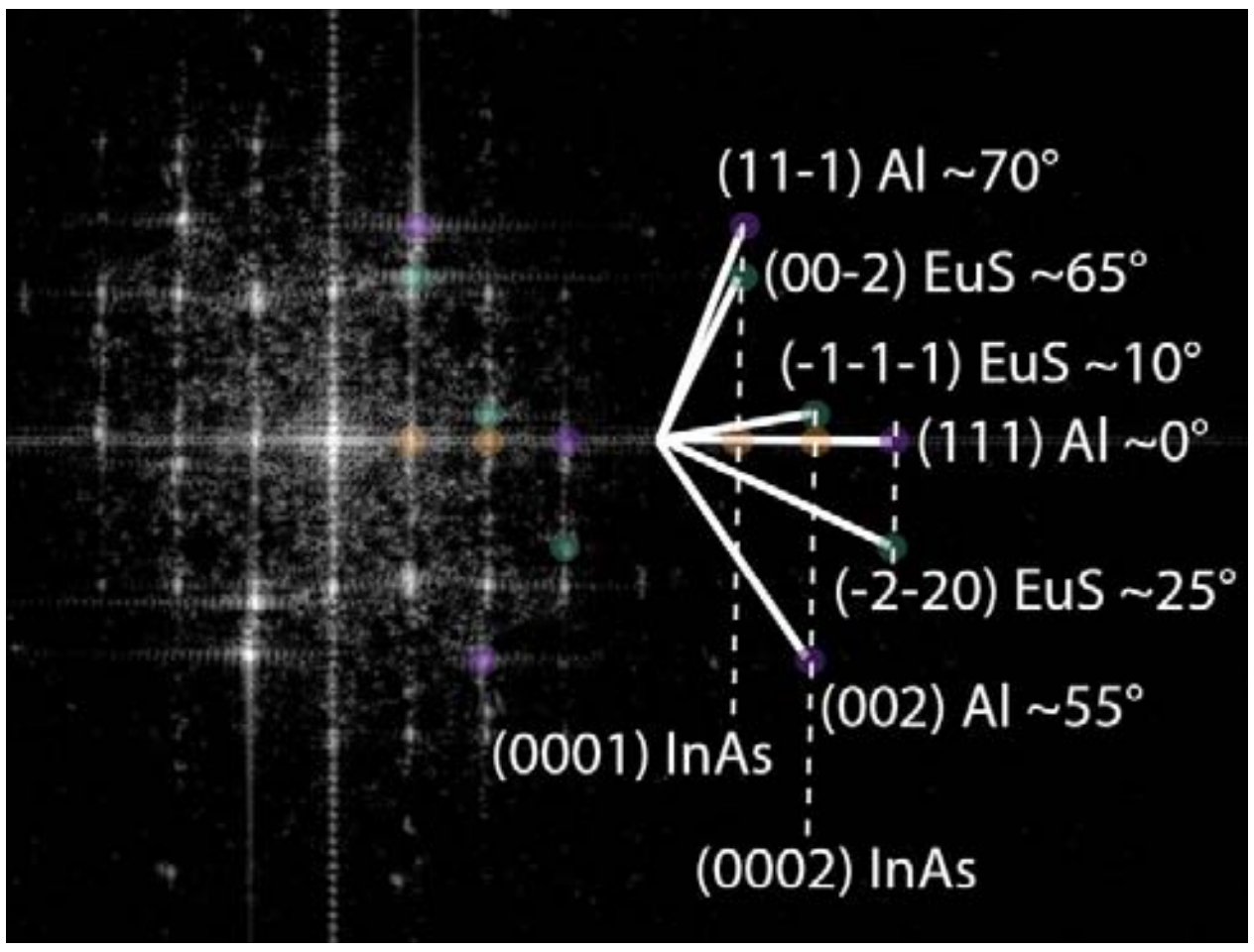

Figure S2 | Indexed power spectrum (FFT) to demonstrate the coherence of lattice between InAs and EuS as well as between EuS and Al with zone axis transverse to the {1-100}-type corner facet of the InAs/EuS/Al NW. Indexes from InAs are in orange and indexes from EuS are colored light green, while Al light purple.

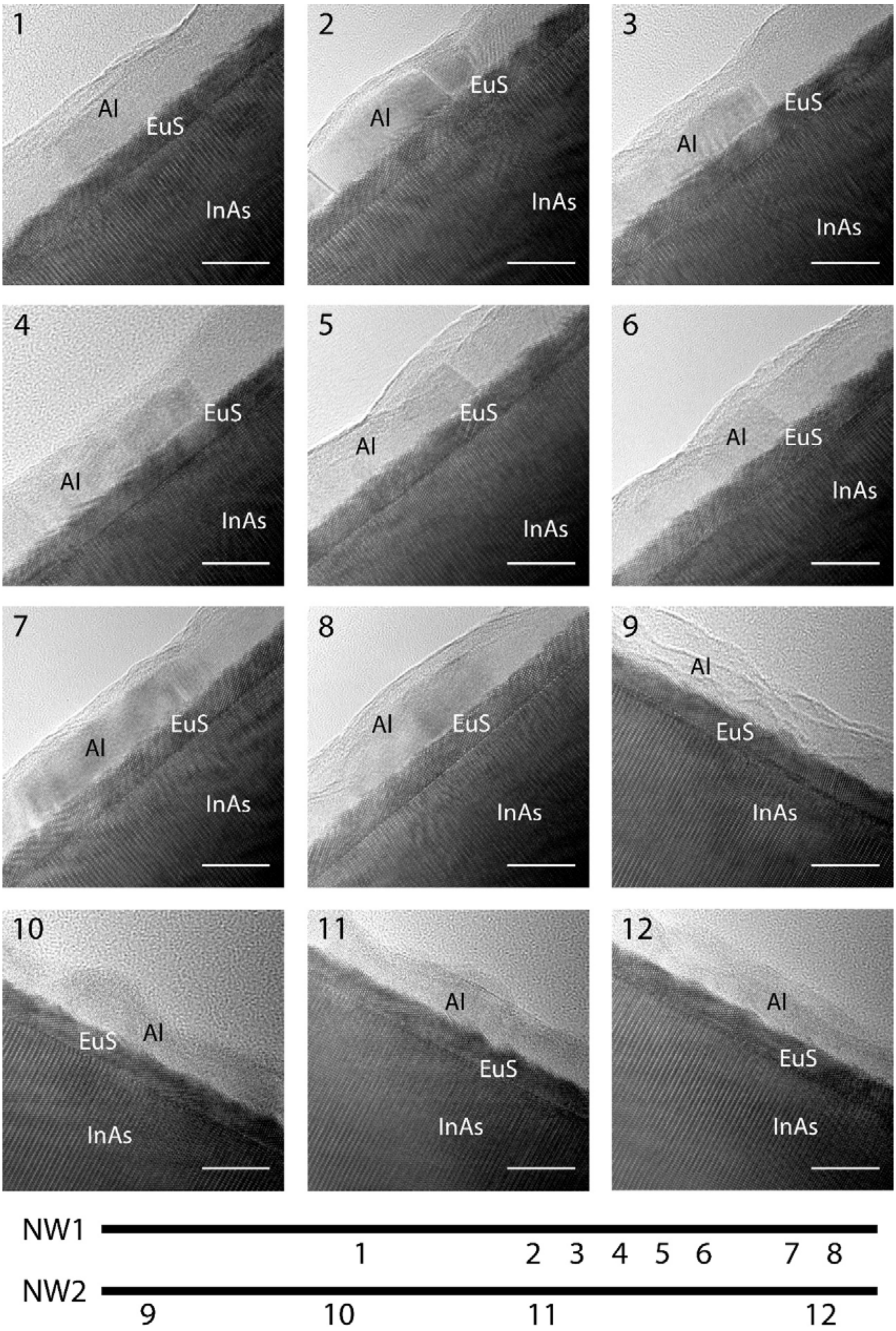

1
Al
EuS
InAs
2
EuS
Al
InAs
3
EuS
Al
InAs
4
EuS
Al
InAs
5
EuS
Al
InAs
6
Al
EuS
InAs
7
EuS
Al
InAs
8
Al
EuS
InAs
9
Al
EuS
InAs
10
EuS
Al
InAs
11
Al
EuS
InAs
12
Al
EuS
InAs
NW1
1 2 3 4 5 6 7 8
NW2
9 10 11 12

Figure S3 | HRTEM images obtained with zone axis transverse to the {1-100}-type corner facet of the InAs/EuS/Al NW in the different sections of two NWs. The positions of images are schematic at the bottom. The scale bar shows 10 nm.

**2. magnetic properties of InAs/EuS/Al VLS NWs**

2.1 scanning SQUID of InAs/EuS/Al VLS NWs

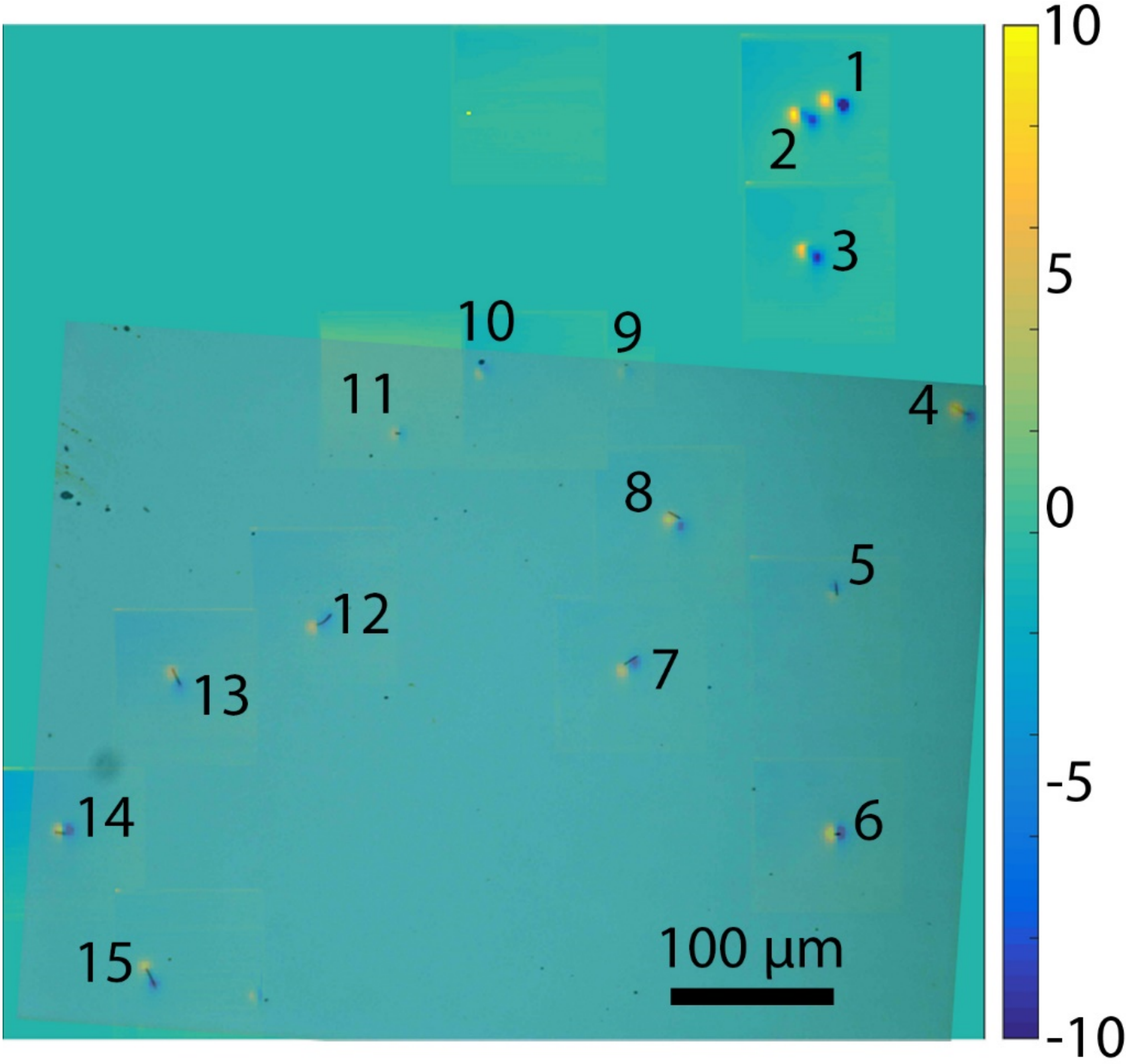

Figure S4 | The SQUID mapping on the area having InAs / EuS NWs. The one-to-one relation is confirmed between the measured dipoles and the NWs. The No. 1-3 NWs are not shown in order to get good scanning contrast.

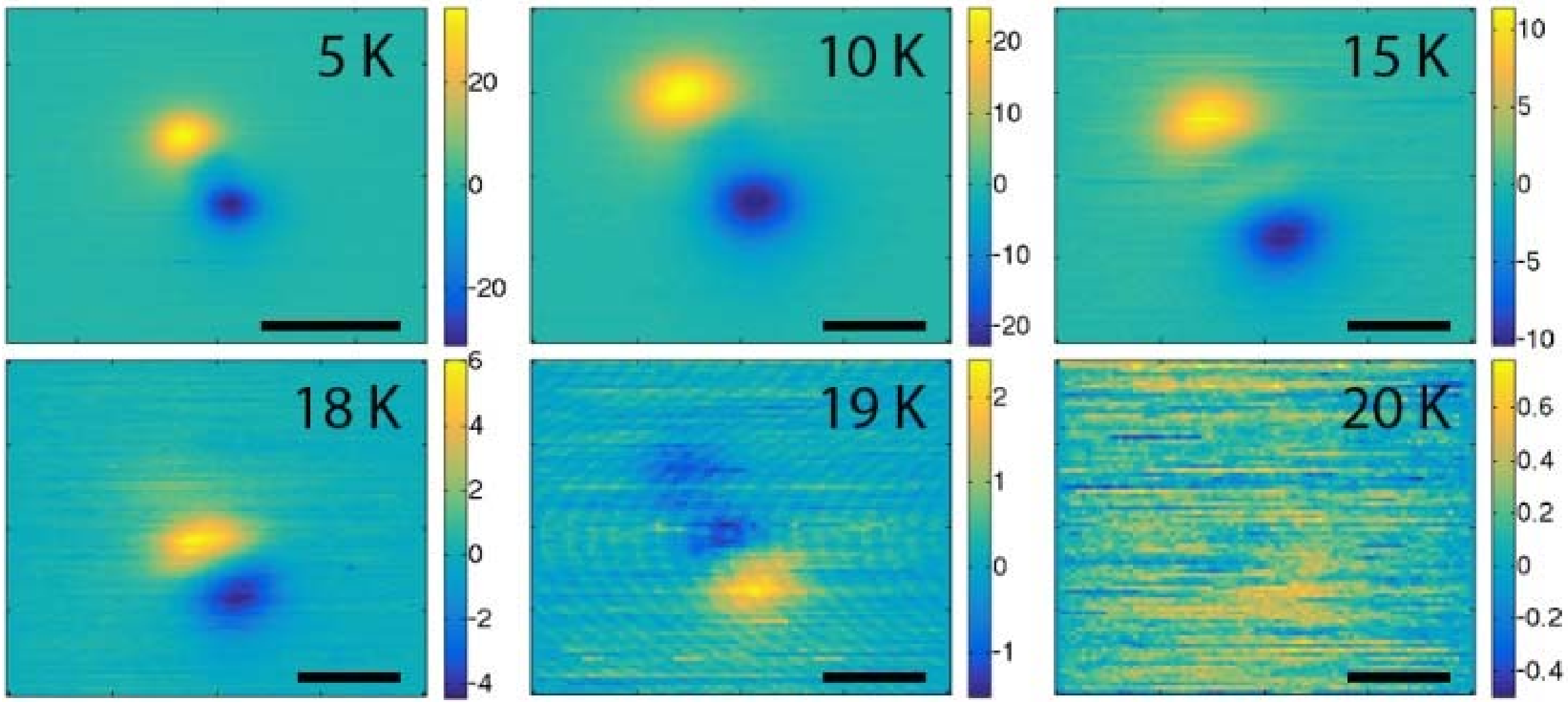

Figure S5 | The SQUID measurement of a 10 µm InAs / EuS NW warmed up from 5 K to 20 K showing the magnetic phase transition and the corresponding Curie temperature of 19 K. The NW was cooled to 5 K with a 65 Oe field before warming up. The scale bars show 10 µm.

Table S1 | The peak-to-peak magnetometry signal of each NW measured by scanning SQUID when cooled in zero field or with 65 Oe and corresponding average values. The number of NWs is consistent with that in Fig. S4.

| NW number | Peak-to-peak magnetometry signal when cooled in zero B field ($m\Phi_0$) | Peak-to-peak magnetometry signal when cooled with 65 Gauss ($m\Phi_0$) |
|---|---|---|
| 1 | 36 | 96 |
| 2 | 16 | 82 |
| 3 | 19 | 60 |
| 4 | 52 | 58 |
| 5 | 4 | 29 |
| 6 | 51 | 178 |
| 7 | 23 | 63 |
| 8 | 55 | 58 |
| 9 | 2 | 14 |
| 10 | 17 | 40 |
| 11 | 3 | 33 |
| 12 | 4 | 43 |
| 13 | 35 | 50 |
| 14 | 11 | 90 |
| 15 | 26 | 16 |
| **Average** | **24** | **61** |

2.2 Magnetic field simulation of InAs/EuS/Al VLS NWs without superconductivity

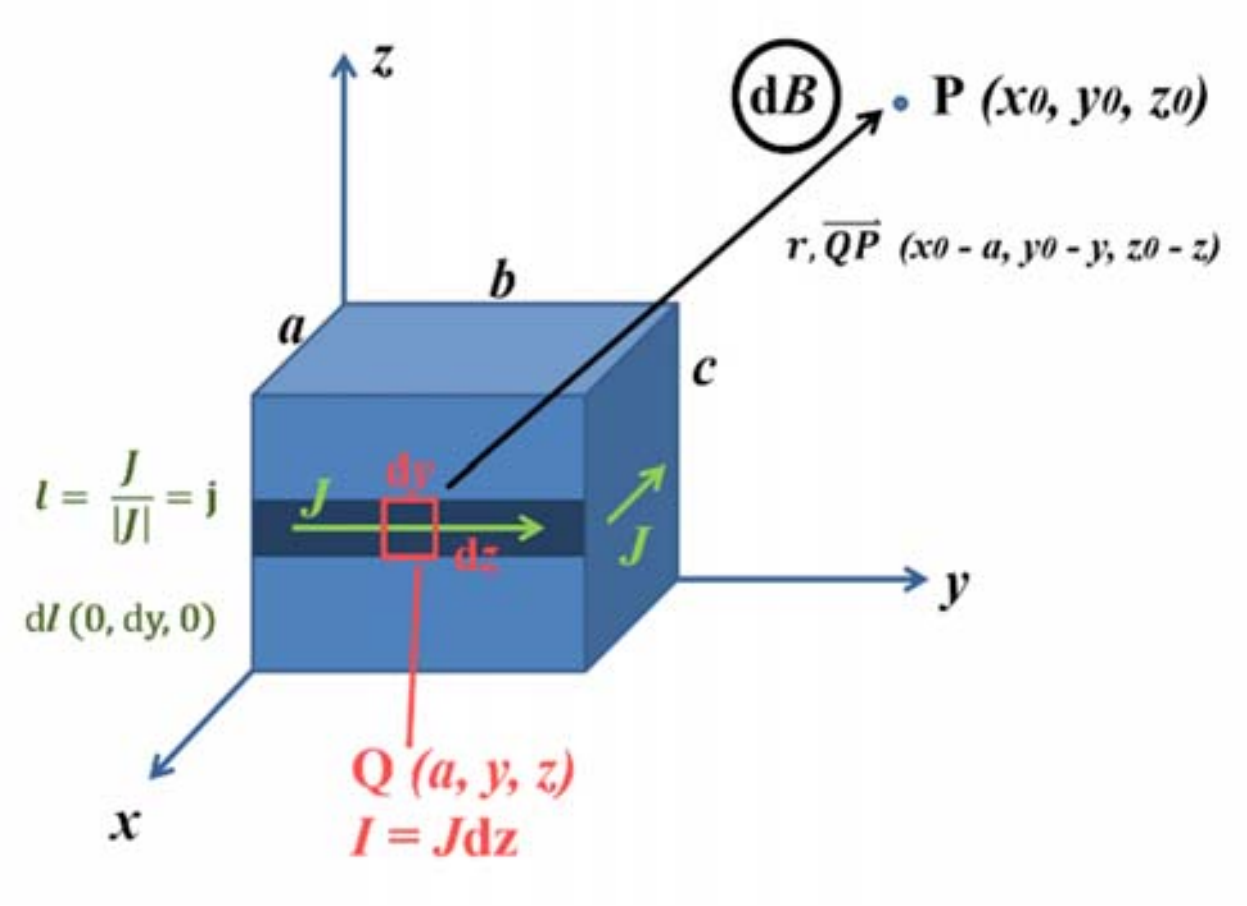

Figure S6 | The schematic of the model for the magnetic field simulation of a cuboid magnet.

In this case, we have assumptions as follows:

1. The magnets follow the Ampère's electrodynamic molecular model.
2. The magnetic permeability of both InAs and Al is equal to the vacuum magnetic permeability as InAs is diamagnetic and Al is also diamagnetic when it is not superconducting.
3. The epitaxial EuS is perfect cuboid and magnetized evenly along NWs.
4. The magnetization is along NWs.

As shown in Fig. S6, the cuboid magnet has the size of a × b × c. We would like to know the magnetic field at the point P $(x_0, y_0, z_0)$. The uniform magnetization will bring the same current in each molecule. The field is excited by all molecules while the current with the same magnitude and the opposite directions will neutralize each other, so the current will only exist on the surface and of course it is constant. According to the Biot−Savart Law, the magnetic field of a random point Q can be described as:

$$\mathrm{d}\boldsymbol{B} = \frac{\mu\mu_0}{4\pi}\frac{I\mathrm{d}\boldsymbol{l}\times\boldsymbol{r}}{r^3} \tag{1}$$

Here, $\mathrm{d}\boldsymbol{B} = \mathrm{d}B_x\boldsymbol{i} + \mathrm{d}B_y\boldsymbol{j} + \mathrm{d}B_z\boldsymbol{k}$ is the field from the point, $\mu$ is the relative magnetic permeability and set to 1 as InAs, $\mu_0 = 4\pi \times 10^{-7}\ \mathrm{J/(A^2m)}$ is the magnetic constant, $I = J\mathrm{d}z$ is the molecular current at the point, $\boldsymbol{l} = \frac{\mathrm{J}}{|J|} = \mathrm{j}$ is the victor of the current and $\boldsymbol{r}$ is the victor $\overrightarrow{\boldsymbol{QP}}$. For an example, if the point Q on the surface (a, y, z), $\mathrm{d}\boldsymbol{l} = (0, \mathrm{dy}, 0)$ and $\boldsymbol{r} = \overrightarrow{\boldsymbol{QP}} = (x_0 - a, y_0 - y, z_0 - \mathrm{z})$. Based on this, the contribution of a cross-section plate with the thickness of dz will be:

$$\mathrm{d}B_x = K\mathrm{d}z\int_0^b [\Psi_3(x_0 - a, y_0 - y, z_0 - z) - \Psi_3(x_0, y_0 - y, z_0 - z)]\mathrm{d}y \tag{2}$$

$$\mathrm{d}B_y = K\mathrm{d}z\int_0^a [\Psi_3(x_0 - x, y_0 - b, z_0 - z) - \Psi_3(x_0 - x, y_0, z_0 - z)]\mathrm{d}x \tag{3}$$

$$\mathrm{d}B_z = K\mathrm{d}z\int_0^a [\Psi_2(x_0 - x, y_0, z_0 - z) - \Psi_2(x_0 - x, y_0 - b, z_0 - z)]\mathrm{d}x + K\mathrm{d}z\int_0^b [\Psi_1(x_0, y_0 - y, z_0 - z) - \Psi_1(x_0 - a, y_0 - y, z_0 - z)]\mathrm{d}y \tag{4}$$

Where

$$K = \frac{\mu\mu_0 J}{4\pi} \tag{5}$$

$$J = \frac{I_{total}}{c} = \frac{\frac{M_s}{S}}{c} = \frac{M_s}{abc} = \frac{\frac{M_s}{c}}{ab} = \frac{m_s}{ab} \tag{6}$$

$$\Psi_i(\tau_1, \tau_2, \tau_3) = \frac{\tau_i}{({\tau_1}^2 + {\tau_2}^2 + {\tau_3}^2)^{\frac{3}{2}}} \quad (i = 1,2,3) \tag{7}$$

$M_s$ is magnetization and $m_s$ is magnetization per unit length. It is necessary to convert from the unit of $\mu_B$ to $Am^2$ with 1 $\mu_B = 9.274 \times 10^{-24}$ $Am^2$.

So the field components are:

$$B_x = \int_0^c \mathrm{d}B_x = \frac{K}{2}\begin{pmatrix} \Xi_1(x_0 - a, y_0 - b, z_0 - c) - \Xi_1(x_0 - a, y_0, z_0 - c) \\ +\Xi_1(x_0, y_0, z_0 - c) - \Xi_1(x_0, y_0 - b, z_0 - c) \end{pmatrix} - \frac{K}{2}\begin{pmatrix} \Xi_1(x_0 - a, y_0 - b, z_0) - \Xi_1(x_0 - a, y_0, z_0) \\ +\Xi_1(x_0, y_0, z_0) - \Xi_1(x_0, y_0 - b, z_0) \end{pmatrix} \tag{8}$$

$$B_y = \int_0^c \mathrm{d}B_y = \frac{K}{2}\begin{pmatrix} \Xi_2(x_0 - a, y_0 - b, z_0 - c) - \Xi_2(x_0 - a, y_0, z_0 - c) \\ +\Xi_2(x_0, y_0, z_0 - c) - \Xi_2(x_0, y_0 - b, z_0 - c) \end{pmatrix} - \frac{K}{2}\begin{pmatrix} \Xi_2(x_0 - a, y_0 - b, z_0) - \Xi_2(x_0 - a, y_0, z_0) \\ +\Xi_2(x_0, y_0, z_0) - \Xi_2(x_0, y_0 - b, z_0) \end{pmatrix} \tag{9}$$

$$\begin{aligned} B_z = \int_0^c \mathrm{d}B_z = &-K\begin{pmatrix} \Xi_3(x_0 - a, y_0 - b, z_0 - c) - \Xi_3(x_0 - a, y_0, z_0 - c) \\ +\Xi_3(x_0, y_0, z_0 - c) - \Xi_3(x_0, y_0 - b, z_0 - c) \end{pmatrix} \\ &+ K\begin{pmatrix} \Xi_3(x_0 - a, y_0 - b, z_0) - \Xi_3(x_0 - a, y_0, z_0) \\ +\Xi_3(x_0, y_0, z_0) - \Xi_3(x_0, y_0 - b, z_0) \end{pmatrix} \\ &- K\begin{pmatrix} \Xi_4(x_0 - a, y_0 - b, z_0 - c) - \Xi_4(x_0 - a, y_0, z_0 - c) \\ +\Xi_4(x_0, y_0, z_0 - c) - \Xi_4(x_0, y_0 - b, z_0 - c) \end{pmatrix} \\ &+ K\begin{pmatrix} \Xi_4(x_0 - a, y_0 - b, z_0) - \Xi_4(x_0 - a, y_0, z_0) \\ +\Xi_4(x_0, y_0, z_0) - \Xi_4(x_0, y_0 - b, z_0) \end{pmatrix} \end{aligned} \tag{10}$$

Where

$$\Xi_1(\tau_1,\tau_2,\tau_3) = ln\frac{\sqrt{{\tau_1}^2+{\tau_2}^2+{\tau_3}^2}-\tau_2}{\sqrt{{\tau_1}^2+{\tau_2}^2+{\tau_3}^2}+\tau_2} \tag{11}$$

$$\Xi_2(\tau_2,\tau_1,\tau_3) = ln\frac{\sqrt{{\tau_1}^2+{\tau_2}^2+{\tau_3}^2}-\tau_2}{\sqrt{{\tau_1}^2+{\tau_2}^2+{\tau_3}^2}+\tau_2} \tag{12}$$

$$\Xi_3(\tau_1,\tau_2,\tau_3) = arctan\frac{\tau_2}{\tau_1}\frac{\tau_3}{\sqrt{{\tau_1}^2+{\tau_2}^2+{\tau_3}^2}} \tag{13}$$

$$\Xi_4(\tau_2,\tau_1,\tau_3) = arctan\frac{\tau_2}{\tau_1}\frac{\tau_3}{\sqrt{{\tau_1}^2+{\tau_2}^2+{\tau_3}^2}} \tag{14}$$

Notice that the positions of independent variables $\tau_1$ and $\tau_2$ between the function $\Xi_1$ and $\Xi_2$, between the function $\Xi_3$ and $\Xi_4$ are shifted, while the function forms are kept consistent.

Based on the equations (5), (6), (8)-(14), the field can be decided as long as the size of the magnet (a, b, c), the position of the point in space $(x_0, y_0, z_0)$ and the magnetization $M_s$ (or the magnetization per unit length $m_s$) are known.

The magnitude of $\boldsymbol{B}$ in Fig. S7a is obtained via:

$$|\boldsymbol{B}| = \left|B_x\boldsymbol{i} + B_y\boldsymbol{j} + B_z\boldsymbol{k}\right| = \sqrt{{B_x}^2 + {B_y}^2 + {B_z}^2} \tag{15}$$

And $B_{//}$ in Fig. S7b is the field component along z axis $B_z$.

The parameters to demonstrate simulation in the context:

1. The size of the magnet is based on that of EuS layers, which is 10 μm × 50 nm × 3 nm.
2. The average NW magnetization 3 × $10^7$ $\mu_B$/μm is employed based on estimation of the peak-to-peak magnetometry. It is not reliable because too many assumption will be imported during estimation but could be seen as a reasonable reference.

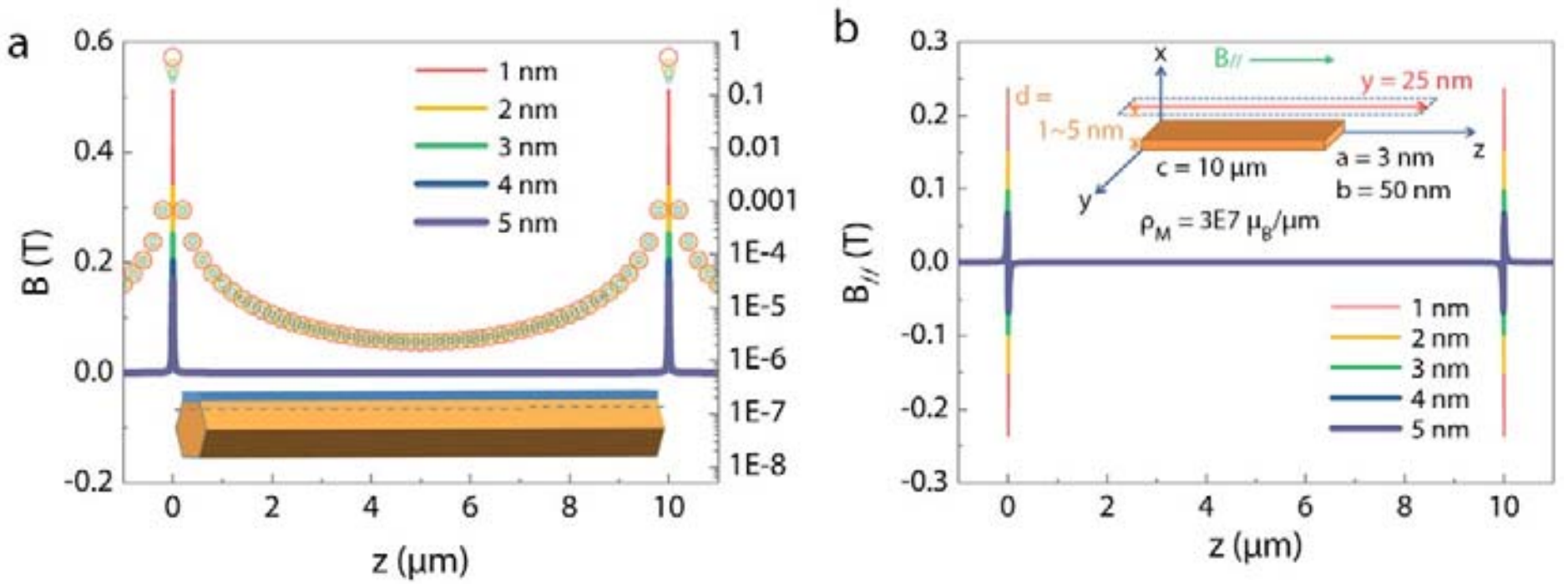

Figure S7 | The magnetic field simulation. a, A simulation of the total magnetic fields of a NW, with the model shown below the curves, which indicates most of the fields at two ends of NWs. b, A simulated magnetic field parallel to [0001] InAs of a NW. The inset provides information about the simulation of magnetic fields.

2.3 The intact EuS layer of InAs/EuS/Al VLS NWs after etching during NW device fabrication

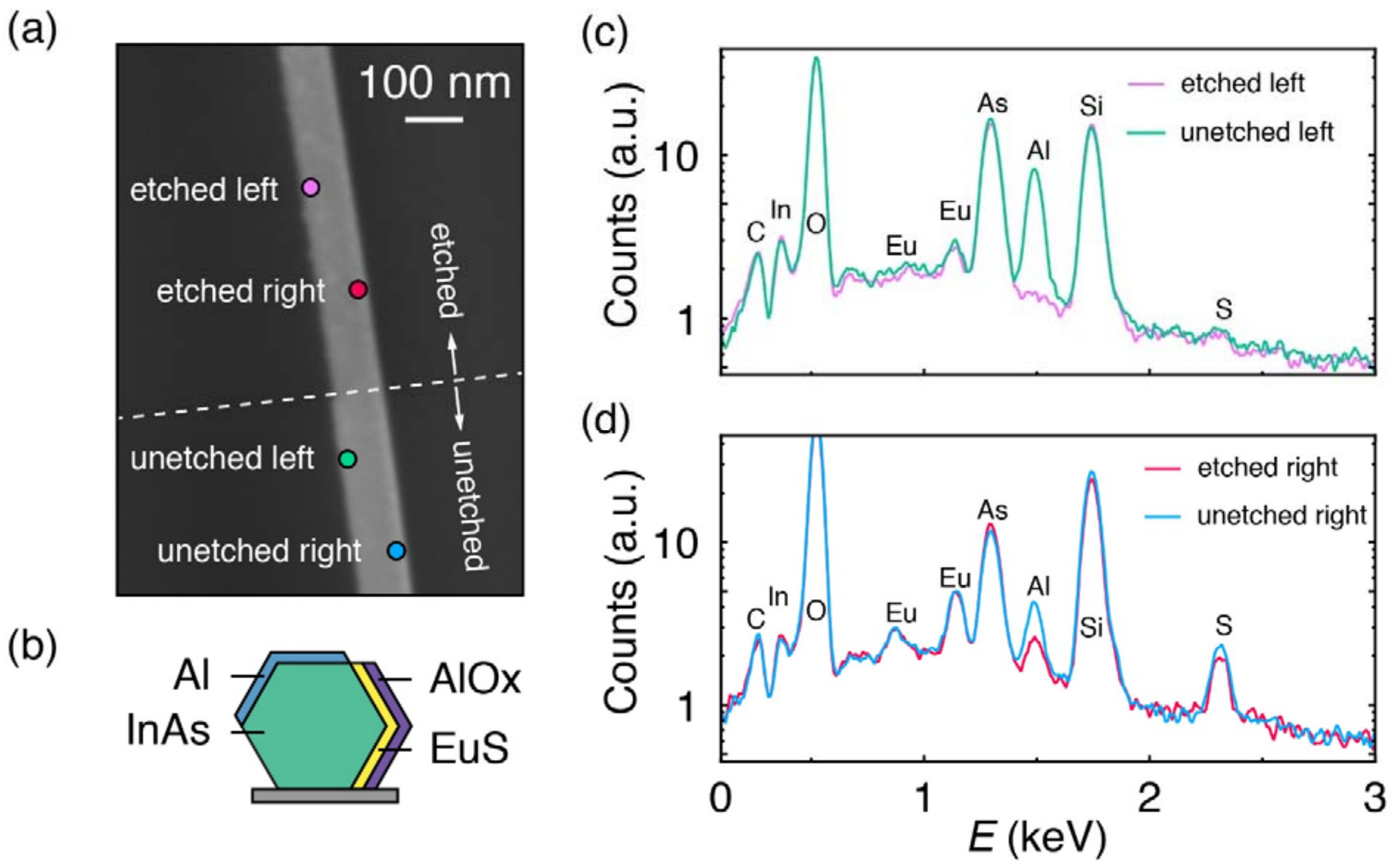

Figure S8 | Energy-dispersive X-ray spectroscopy to show EuS layer intact after etching. a, The SEM image with marks to show where the measurements were carried out. b, Schematic device cross section showing the Al and EuS shells relative to the InAs core. c,d, The comparison of energy-dispersive X-ray spectra showing only Al was etched in the etched area and no EuS was lost during etching.

## 3. Structural information of InAs/EuS SAG NWs

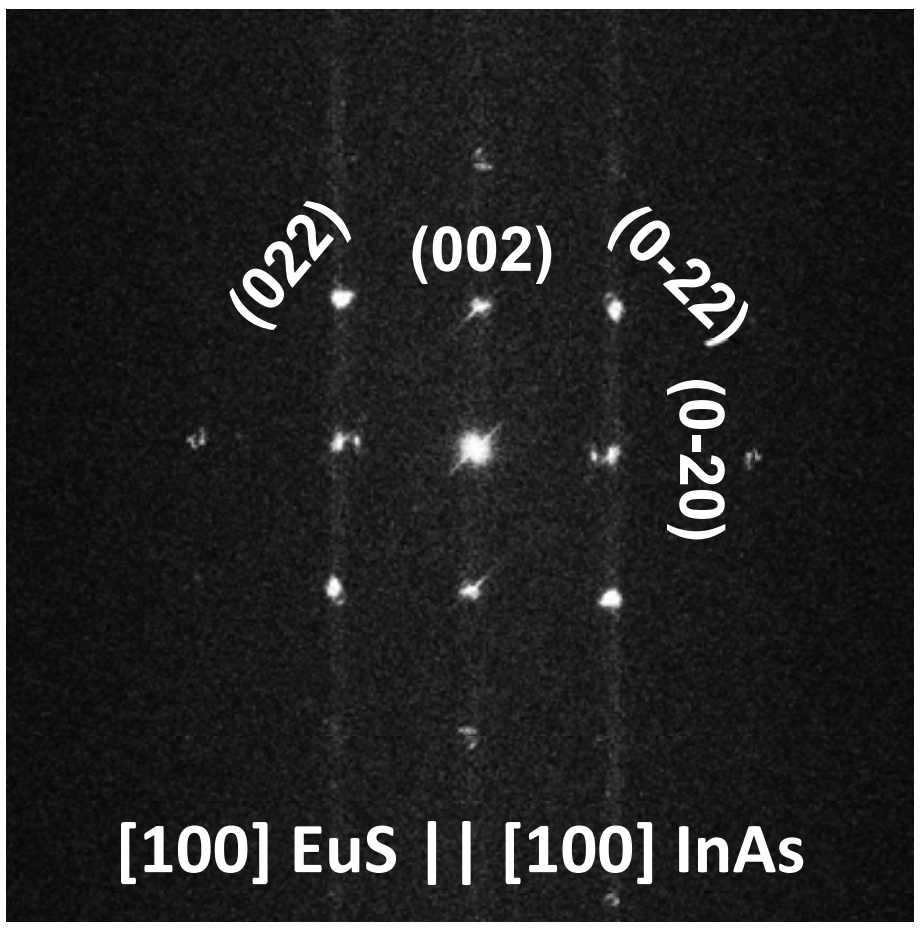

Figure S9 | The indexed power spectrum (FFT) in EuS epitaxial on selective area grown (SAG) InAs.

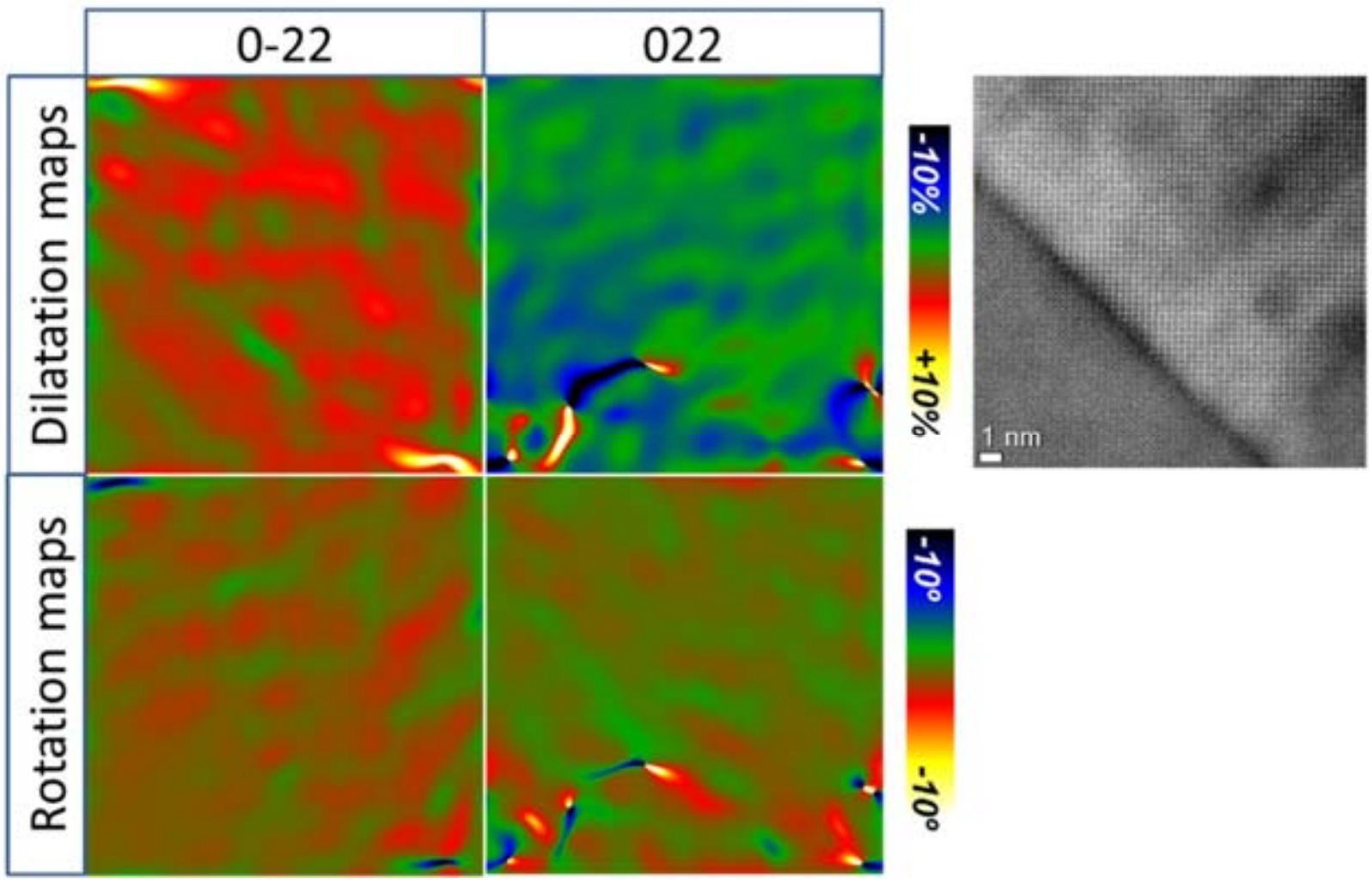

Figure S10 | The GPA dilatation and rotation maps along [0-22] and [022] of a cross-sectional micrograph of a SAG InAs/EuS NW, respectively. The corresponding cross-sectional image is shown on the right panel. Note that artifacts along [022] don’t correspond to the interface so we attribute them to scanning issues.

## 4. scanning SQUID of InAs/EuS SAG NWs

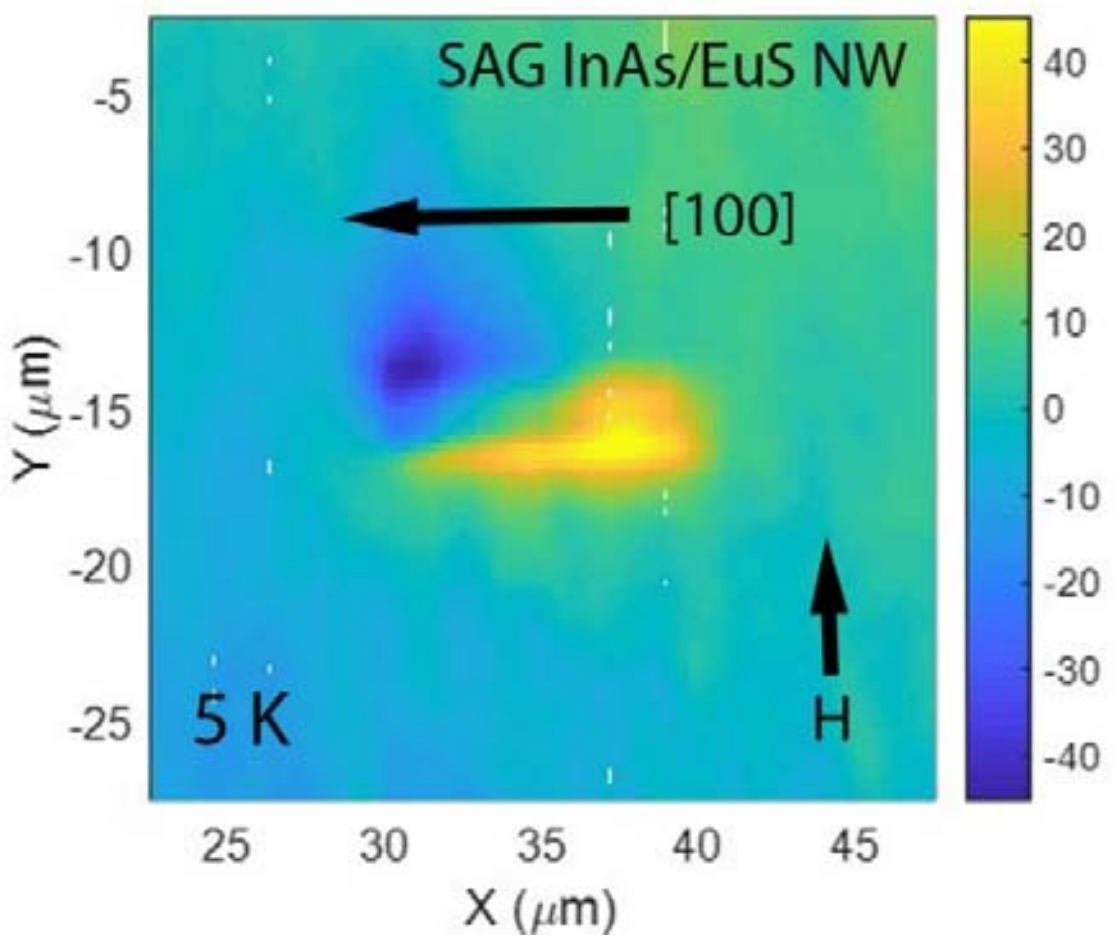

Figure S11 | Scanning SQUID measurement of a SAG InAs/EuS [100] NW at 5 K with in-plane magnetic field cooling perpendicular to [100]. The value of magnetic flux is shown with different colors from yellow to blue.

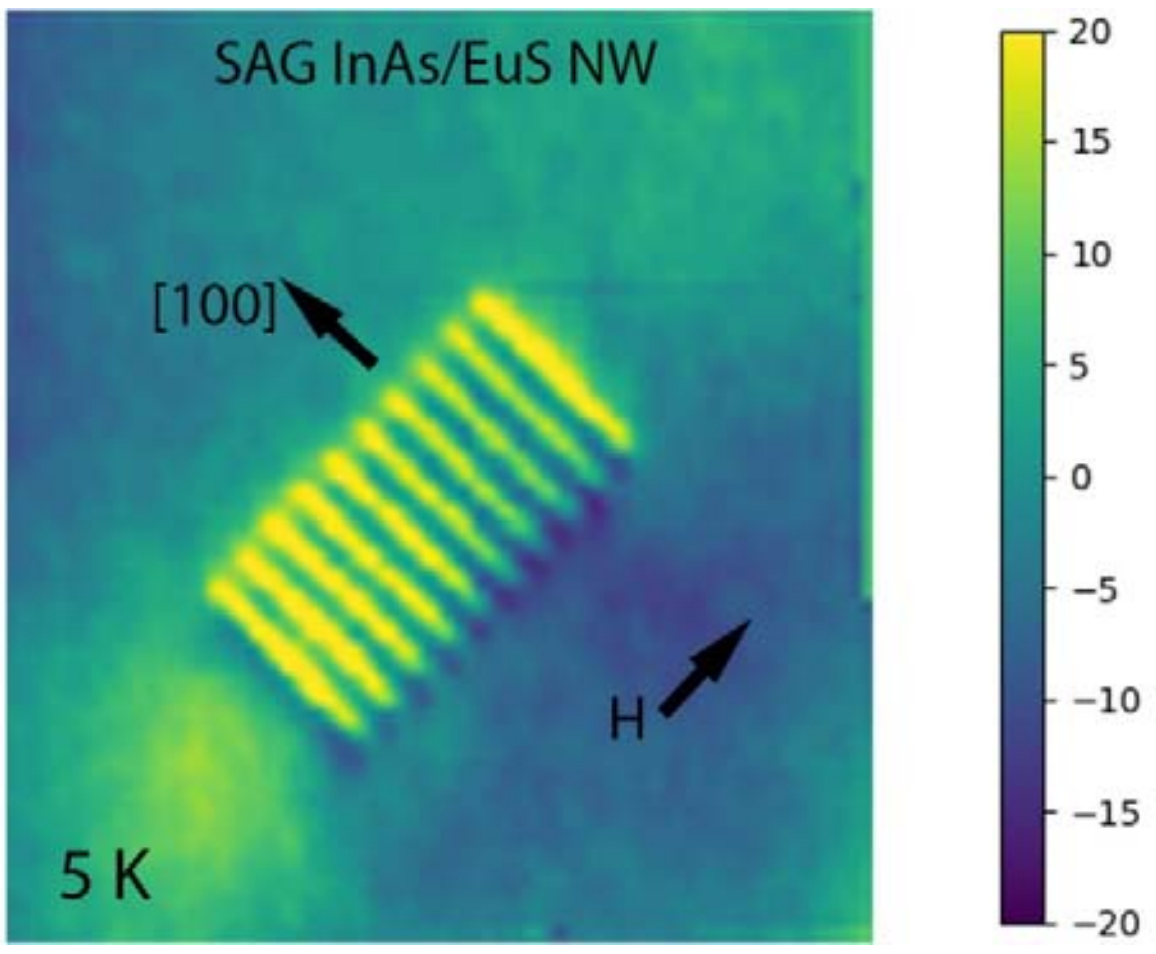

Figure S12 | Scanning SQUID measurement of a SAG InAs/EuS [100] NW array at 5 K with in-plane magnetic field cooling perpendicular to [100]. The value of magnetic flux is shown with different colors from yellow to blue.